\documentclass[12pt, oneside, a4paper]{article}
\usepackage[colorlinks,bookmarksopen]{hyperref}
\usepackage{color}
\usepackage{graphicx}
\usepackage{times}

\begin{document}

\newcommand{\be}{\begin{equation}}
\newcommand{\ee}{\end{equation}}
\newcommand{\bea}{\begin{eqnarray}}
\newcommand{\eea}{\end{eqnarray}}
\newcommand{\f}{\frac}
\newcommand{\p}{\partial}
\newcommand{\no}{\nonumber}
\newcommand{\kB}{k_\mathrm{B}}
\newcommand{\kT}{k_\mathrm{B}T}
\newcommand{\e}{\mathrm{e}}
\newcommand{\dd}{\mathrm{d}}

\newcommand{\eg}{\textit{e.g.}}
\newcommand{\ie}{\textit{i.e.}}
\newcommand{\etc}{\textit{etc}}

\newcommand{\tf}{t_\mathrm{f}}
\newcommand{\Lp}{l_\mathrm{p}}
\newcommand{\LK}{l_\mathrm{K}}
\newcommand{\Laa}{l_\mathrm{aa}}
\newcommand{\Ld}{L_\mathrm{d}}
\newcommand{\Naa}{N_\mathrm{aa}}
\newcommand{\Nd}{N_\mathrm{d}}
\newcommand{\Nu}{N_\mathrm{u}}
\newcommand{\Nb}{N_\mathrm{b}}
\newcommand{\kb}{k_\mathrm{b}}

\newcommand{\vR}{\vec{R}}
\newcommand{\vF}{\vec{F}}
\newcommand{\vL}{\vec{L}}
\newcommand{\vLd}{\vec{L}_\mathrm{d}}

\title{The relevance of neck linker docking in the motility of kinesin}

\author{Andr\'as Cz\"ovek, Gergely J. Sz\"oll\H{o}si, Imre Der\'enyi\\
\textit{\small Department of Biological Physics, E\"otv\"os University}\\
\textit{\small P\'azm\'any P.\ stny.\ 1A, H-1117 Budapest, Hungary}}

\maketitle

\begin{abstract}
Conventional kinesin is a motor protein, which is able to walk along a
microtubule processively. The exact mechanism of the stepping motion
and force generation of kinesin is still far from clear. In this paper
we argue that neck linker docking is a crucial element of this
mechanism, without which the experimentally observed dwell times of the
steps could not be explained under a wide range of loading forces. We
also show that the experimental data impose very strict constraints on
the lengths of both the neck linker and its docking section, which are
compatible with the known structure of kinesin.
%
\end{abstract}

\section{Introduction}

Kinesin motor proteins execute a variety of intracellular transport
functions by transporting cellular cargo along microtubules (MTs) while
hydrolyzing adenosine triphosphate (ATP)
\cite{Howard_Hudspeth_Vale,Hackney,Gelles,vale_evo}.
These molecular walking machines move in 8-nm steps toward the plus end
of microtubules, turning over one ATP molecule per step under a range
of loads
\cite{Rice_Vale,vale,Yidliz_Tomishige_Vale}.
Conventional kinesin (kinesin-1) is a homodimer, the monomers of which 
consist of a head domain (containing a conserved catalytic core), a
stalk (through which the monomers form a dimer), and an approximately
13 amino acid long neck linker (connecting the head to the stalk).
The neck linker is evolutionarily highly conserved among
plus-end directed motors
\cite{vale_evo}
and appears to be crucial for motility
\cite{rice2}.

In a kinesin dimer the two stalks form a coiled coil, to which a cargo
is attached or, during experiments, a pulling force is applied. The
catalytic core of each head is responsible for binding and hydrolyzing
ATP
\cite{ATPase,ATPase2,confchange}.
ATP binding to a MT bound head has been shown to result in a section of
the neck linker binding to the head
\cite{rice,rice2,confchange,confchange2},
with a consequence of positioning the remaining unbound section (and
also the diffusing other head) closer to the forward binding site
\cite{Rice_Vale,fox}.
Experimental studies
\cite{rice}
of this conformational change (referred to as `neck linker docking')
strongly support a scenario wherein an unstructured, random-coil-like
neck linker folds onto the core as a result of nucleotide binding --
paying a large entropic cost compensated by a large enthalpic gain
(both in the order of $50$~kJ/mol or $20$~$\kT$, where
$\kB\approx1.38\times10^{-23}$~J/K is the Boltzmann constant and $T$
denotes the absolute temperature, which we set now to be $293$~K).

\begin{figure}[t!]
\centerline{\includegraphics[width=0.9\columnwidth]{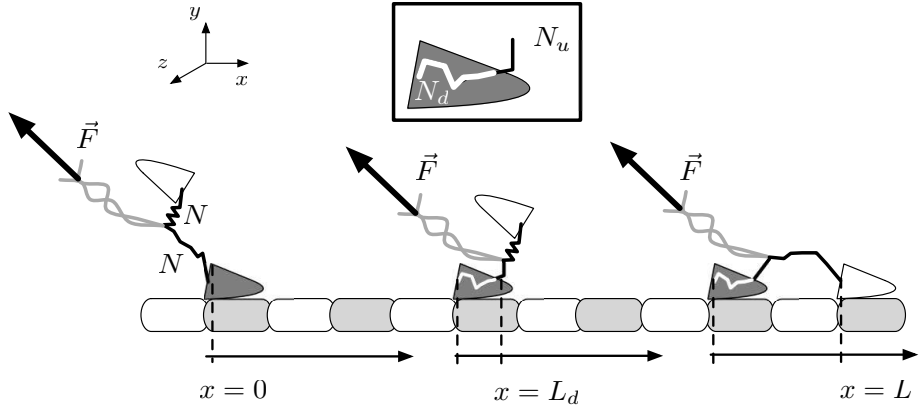}}
\caption{
One head bound kinesin before (left) and after (middle) neck linker
docking, followed by the binding of the tethers head to the
microtubule (right). Inset shows the number of docked neck linker
segments ($\Nd$) in white and that of the remaining ($\Nu$) segments of
the same neck linker in black.
}
\label{fig:kin}
\end{figure}

Neck linker docking is also confirmed by X-ray structures
\cite{Xray1,Xray2},
molecular dynamics simulations
\cite{MD1,MD2},
and fluorescence experiments
\cite{Tomishige},
however, it is still debated whether docking is crucial to the
processivity and force generation of kinesin, or just a byproduct of
some other mechanism
\cite{Guydoshblock}.
To address this question we focus on a single state of the dimer, in
which one of the heads is bound to the MT and contains an ATP, while
the other head (often called the `tethered head') is unbound and
contains an ADP (see left cartoon in Fig.\ \ref{fig:kin}). To complete
a forward step, the tethered head must find and then bind to the
forward binding site on the MT (which is about $L=8$~nm ahead of the
bound head).

In a recent optical tweezers experiment Carter and Cross 
\cite{focikk}
demonstrated that kinesin can take not only forward but also backward
steps. The ratio of the number of backward and forward steps increases
as the loading force is increased and reaches unity at the stall force
of about $-7$~pN (a negative force means pulling toward the minus end
of the MT). At saturating ATP concentration and at zero loading force
the steps almost exclusively occur in the forward direction and their
duration (which is also called the dwell time) is around $0.01$~s. Near
the stall force, however, the steps are much slower with a dwell time
of about $0.3$~s. We note here that in any situation when two
alternative processes compete with each other (such as the forward and
backward steps of kinesin) the duration of the successful process will
be determined by that of the faster one (meaning that the completion of
the slower one will take a similar amount of time, but will occur less
frequently). So at small loads the forward steps, whereas at loads
exceeding the stall force the backward steps will determine the dwell
times of the steps in both directions.

The benefit of neck linker docking appears to be clear: in the one head
bound state (described above) this conformational change positions the
tethered head closer to the forward binding site on the MT (as
demonstrated in Fig.\ \ref{fig:kin}). However, to underpin the
necessity of neck linker docking in a quantitative manner, we examine
under what conditions can the tethered head find the binding site
within the aforementioned $0.3$~s time limit when the $-7$~pN stall
force is applied.

\section{The Model}

We suppose that the undocked sections of the neck linkers behave as
random coils, which we simply model as freely jointed chains (FJCs).
Thus, before binding to the MT the tethered head experiences a random
walk in the potential of the entropic springs formed by the two
connected neck linkers of kinesin. We regard the spatial distribution
of the tethered head simply as that of the end of the two joined neck
linkers. Each neck linker consists of $\Naa$ amino acids ($\Naa \approx
13$ for a conventional kinesin). We use $\Laa=0.38$~nm for one amino
acid length and $\Lp=0.44$~nm for the persistence length of a
polypeptide chain
\cite{persistence}.
The Kuhn length is then $\LK=2\Lp=0.88$~nm, and the number of freely
jointed segments in a neck linker is $N=\Naa\Laa/\LK$, where $\Naa\Laa$
is the contour length of the polypeptide chain
\cite{howard}.
As one can see this description is quite coarse with respect to the
number of amino acids of the neck linker: increasing the number of the
segments $N$ by $1$ is equivalent to increasing the number of the amino
acids $\Naa$ by $\LK/\Laa \approx 2.3$.

For technical reasons we divide the neck linker of the bound head into
two sections. The first one consists of those $\Nd$ segments that can
dock to the head domain, while the second one contains the remaining
$\Nu=N-\Nd$ segments of the neck linker. The external pulling force
$\vF$ acts at the end point of this neck linker, where it joins to the
neck linker of the tethered head (as illustrated in Fig.\
\ref{fig:kin}). For simplicity we neglect the width and flexibility of
the coiled coil stalk between the two neck linkers. This omission
could, however, be largely compensated by a slight increase in the
length of the neck linkers.

Let $x$, $y$, and $z$ be Cartesian coordinates so that the $x$ axis is
parallel to the MT, and the $x$ and $y$ axes span the plane of the
external force $\vF$. This way $\vF$ has no $z$ component. The angle of
the force (\ie, that of the coiled coil stalk) to the MT depends on the
details of the experimental setup, in particular, on the length of the
stalk and the size of the bead in the optical trap. Throughout the
paper we use a reasonable value of $45^\circ$ for this angle and, thus,
assume that $F_{y}=|F_{x}|$. In the experiments, usually the $x$
component of $\vF$ are reported, so by specifying that the stall force
is $-7$~pN we mean $\vF=(-7,7,0)$~pN.

Let $\rho_{N}^{0}(\vR)$ denote the probability density of the
end-to-end vector $\vR$ of a free ($\vF=0$) neck linker with $N$
segments. Such a distribution applies to the neck linker of the
tethered head. As $\rho_{N}^{0}(\vR)$ does not depend on the external
force it has a spherical symmetry. The formula for this distribution is
derived in the Appendix.

Let $\rho_{N}(\vR,\vF)$ denote the probability density of the
end-to-end vector $\vR$ (pointing from the head end of the neck linker
towards the pulled end) of the MT-bound head's neck linker with $N$
segments at an applied external force $\vF$. Note that the applied
force breaks the spherical symmetry of the distribution.
$\rho_{N}(\vR,\vF)$ can be expressed with the help of
$\rho_{N}^{0}(\vR)$ as
\begin{equation}
\rho_{N}(\vR,\vF) =
 \frac{\rho_{N}^{0}(\vR) \e^{\frac{\vF\vR}{\kT}}}{Z_{N}(\vF)}
\label{eq:rhob}
\end{equation}
where
\begin{equation}
Z_{N}(\vF) = \int 
 \rho_{N}^{0}(\vR) \e^{\frac{\vF\vR}{\kT}} \dd\vR
\label{eq:z1}
\end{equation}
is the partition function. 

The compound probability density $\rho_{N_1,N_2}(\vR,\vF)$ for the
end-to-end vector $\vR$ of two joint neck linkers (with $N_1$ and $N_2$
segments, respectively) that are pulled by an external force $\vF$ at
the joint (and held fixed at the other end point of the first neck
linker) can then be expressed as:
\begin{equation}
\rho_{N_1,N_2}(\vR,\vF) =
 \int \rho_{N_1}(\vR',\vF) \rho_{N_2}^{0}(\vR-\vR') \dd\vR'
\, .
\label{eq:conv}
\end{equation}

By placing the origin of the coordinate system to the starting point of
the neck linker of the bound head, the probability density
$\rho_{N,N}(\vR,\vF)$ can be considered as the concentration of the
tethered head at position $\vR$, given that the neck linker of the
bound head is undocked, and if all steric constraints are neglected.

To take the volume exclusion between the tethered head and both the MT
(approximated as $R_y<0$) and the bound head (approximated as
$|\vR|<2$~nm) into account, we introduce a constraining function
\be
\Theta(\vR) = \left\{ \begin{array}{ll}
 0 & \textrm{if } R_y<0 \textrm{ or } |\vR|<2 \textrm{~nm,}\\
 1 & \textrm{otherwise.}
\end{array} \right.
\label{eq:theta}
\ee
Multiplying $\rho_{N,N}(\vR,\vF)$ by $\Theta(\vR)$, and then
renormalizing it to unity results in the sterically constrained
concentration of the tethered head:
\be
c(\vR,\vF) =
 \frac{\rho_{N,N}(\vR,\vF) \Theta(\vR)}
 {\int \rho_{N,N}(\vR,\vF) \Theta(\vR) \dd\vR}
\, .
\label{eq:c}
\ee
Note, however, that the volume exclusion is not imposed on the entire
chain, only on its end point. Nevertheless, this steric constraint has
very little effect on our main conclusions.

Similarly, denoting the end-to-end vector of the $\Nd$ segments of the
docked section of the neck linker by $\vLd=(\Ld,0,0)$, \ie, assuming
that docking occurs in the $x$ direction with a projected distance of
$\Ld$, the probability density $\rho_{\Nu,N}(\vR-\vLd,\vF)$ can be
considered as the concentration of the tethered head at position $\vR$
when that the neck linker of the bound head is docked (and all steric
constraints are neglected). After its multiplication by $\Theta(\vR)$
and renormalization to unity, results in the sterically constrained
concentration
\be
c^{*}(\vR,\vF) =
 \frac{\rho_{\Nu,N}(\vR-\vLd,\vF) \Theta(\vR)}
 {\int \rho_{\Nu,N}(\vR-\vLd,\vF) \Theta(\vR) \dd\vR}
\, .
\label{eq:cc}
\ee

Multiplying these concentrations at the positions $\vR=\vL=(L,0,0)$ and
$\vR=-\vL=(-L,0,0)$ by the binding rate constant $\kb$ of the kinesin
head, we get the rate at which the tethered head binds to the MT at the
forward and backward binding sites, respectively.

The only remaining quantity to determine is the probability of the neck
linker being docked or undocked (disregarding the possibility of any
long lived partially docked conformation). These probabilities must be
a function of the force since the neck linker docks easier if it is
pulled forward. Let $P^{*}(\vF)$ denote the probability that the neck
linker is docked and $P(\vF)=1-P^{*}(\vF)$ that it is not. Then the
free energy difference $\Delta G(\vF)$ between the docked and undocked
state can be defined through
\begin{equation}
\frac{P^{*}(\vF)}{P(\vF)} = \e^{\frac{-\Delta G(\vF)}{\kT}}
\, .
\label{eq:neckdock}
\end{equation}

From the measurements of Rice et al.
\cite{rice}
we know that if an ATP is bound to the head, then neck linker docking
is an energetically favorable process with a free energy difference of
$\Delta G^{0} \approx -2 \kT$. Thus, the ratio of the probabilities
that the bound head is in the docked and the undocked conformation can
be written as
\be
\frac{P^{*}(\vF)}{P(\vF)} =
 \frac{\e^{\frac{-\Delta G^{0}}{\kT}}
 \int\int
 \rho_{\Nu}^{0}(\vR'-\vLd) \e^{\frac{\vF\vR'}{\kT}}
 \rho_{N}^{0}(\vR-\vR') \Theta(\vR)
 \dd\vR' \dd\vR}
 {\int\int
 \rho_{N}^{0}(\vR') \e^{\frac{\vF\vR'}{\kT}}
 \rho_{N}^{0}(\vR-\vR') \Theta(\vR)
 \dd\vR' \dd\vR}
\, .
\label{eq:ppp}
\ee

The characteristic time for binding forward at force $\vF$ is then:
\begin{equation}
\tf(\vF) =
 \frac{1}
   {\kb [P^{*}(\vF)c^{*}(\vL,\vF)+P(\vF)c(\vL,\vF)]}
\, .
\label{eq:tf}
\end{equation}

We have no reason to assume a different rate constant $\kb$ for
backward and forward binding, therefore, the ratio of the probabilities
of forward and backward stepping is:
\begin{equation}
r(\vF) =
 \frac{P^{*}(\vF)c^{*}(\vL,\vF)+P(\vF)c(\vL,\vF)}
      {P^{*}(\vF)c^{*}(-\vL,\vF)+P(\vF)c(-\vL,\vF)}
\, .
\label{eq:r}
\end{equation}

\begin{figure}[t!]
\centerline{\includegraphics[width=1.0\columnwidth]{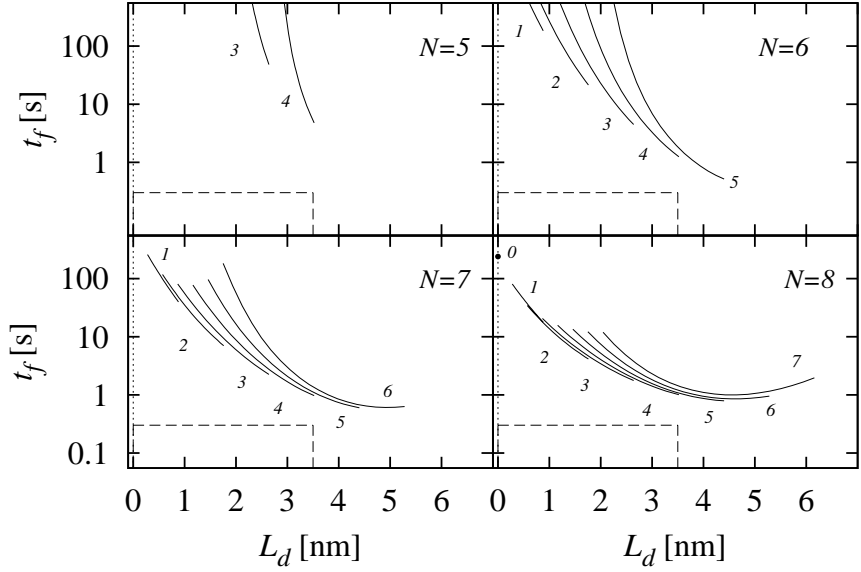}}
\caption{
The characteristic time $\tf$ needed to find the forward binding site
as a function of the horizontal projection $\Ld$ of the docked neck
linker at the $-7$~pN stall force. The four boxes correspond to neck
linkers of different lengths with different numbers $N$ of Kuhn
segments. The various lines in the boxes correspond to different
numbers $\Nd$ of docked segments. The rectangular area in the bottom
left corner of each box indicates the expected range $\tf<0.3$~s and
$\Ld<3.5$~nm, compatible with the experiments.
}\label{fig:tf}
\end{figure}

\begin{figure}[t!]
\centerline{\includegraphics[width=1.0\columnwidth]{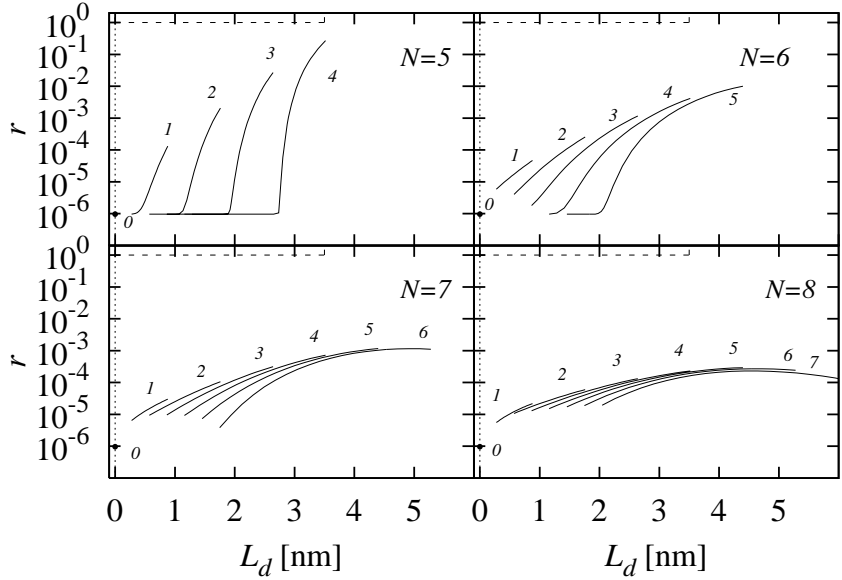}}
\caption{
The ratio $r$ of probabilities to bind forward and backward at the
$-7$~pN stall force as a function of the horizontal projection $\Ld$ of
the docked neck linker. The four boxes correspond to different values
of $N$, and the different lines in the boxes correspond to different
numbers $\Nd$ of docked segments. The preferred range of $r>1$ and
$\Ld<3.5$~nm can be seen in the top left corner of each box.
}\label{fig:r}
\end{figure}

\section{Results}

We evaluated the integrals numerically for four integer values of $N$
between $5$ and $8$. $\Nd$ took values between $0$ and $N-1$, so that
we could compare the situation $\Nd=0$, where neck linker docking plays
no role in the searching of the binding site, to various neck linker
docking geometries. Although for completeness we varied the horizontal
projection $\Ld$ of the docked neck linker from $0.4\Nd\LK$ to
$\Nd\LK$, only values below $3.5$~nm are acceptable since the kinesin
head has a diameter of around $4$~nm
\cite{kinesinsize}.

The results for the characteristic forward binding time $\tf$ at the
$-7$~pN stall force are presented in Fig.\ \ref{fig:tf}. All the curves
decrease monotonically on the interval $\Ld\in[0,3.5\mathrm{nm}]$.
Thus, if we want to find the fastest forward step it is enough to look
at the curves at $\Ld=3.5$~nm. The rectangular areas in the bottom left
corners of the four boxes indicate the desired range of $\tf<0.3$~s and
the allowed regime of $\Ld<3.5$~nm.

\begin{figure}[t!]
\centerline{\includegraphics[width=1.0\columnwidth]{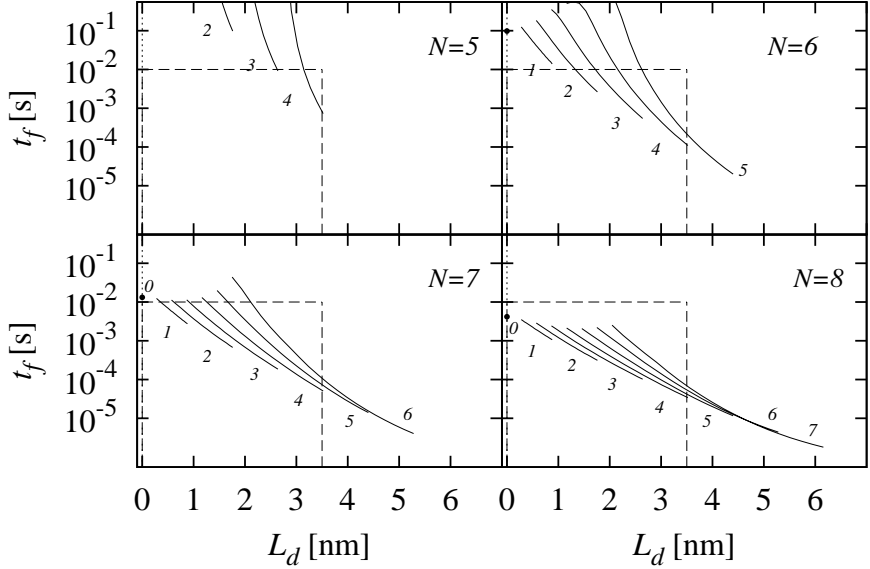}}
\caption{
The same as Fig.\ \ref{fig:tf}, but at zero load, and the
experimentally acceptable range is $\tf<0.01$~s.
}\label{fig:tf0}
\end{figure}

\begin{figure}[t!]
\centerline{\includegraphics[width=1.0\columnwidth]{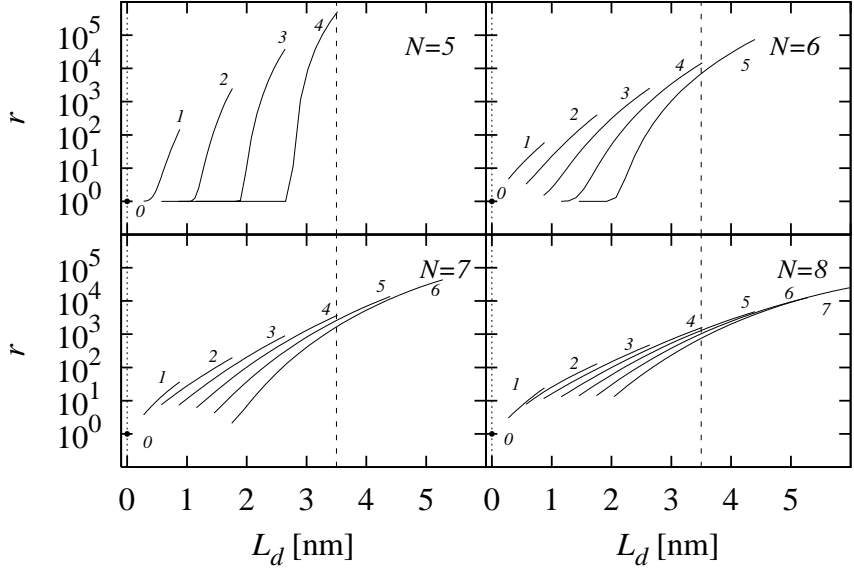}}
\caption{
The same as Fig.\ \ref{fig:r}, but at zero load.
}\label{fig:r0}
\end{figure}

We used $\kb=20$~s$^{-1}$$\mu$M$^{-1}$ for the rate constant of MT
binding
\cite{ratak,MKrate}.
None of the curves passes through the desired area. We can make them do
so by increasing $\kb$ (which would simply shift them downwards). This
can be justified by noticing that the
$\kb\approx20$~s$^{-1}$$\mu$M$^{-1}$ value was measured for free
kinesins in solution. However, in our case the tethered head is
diffusing in the close vicinity of the MT, where electric fields are
not fully screened and local attractive interactions might be
significant, which can lead to an effectively elevated value of $\kb$.
A threefold increase seems sufficient to make some curves just pass
through the desired area for $N\geq6$. A less feasible tenfold increase
would, however, be necessary for $N=5$.

In the $r$ vs. $\Ld$ plots (Fig.\ \ref{fig:r}) the value of $\kb$ does
not play any role (see Eq.\ (\ref{eq:r})). All the curves for $N\geq6$
are orders of magnitudes below $r=1$, the value, which would be the
expected at the stall force, if each forward binding resulted in a
forward step, and each backward binding in a backward step. Obviously,
there must exist some rectification mechanism that makes the completion
of a backward steps less likely (\eg, by reducing the ADP dissociation
rate from the rear head) and increases the value of $r$ to $1$, but
this is not the subject of our investigation.

For comparison we repeated the calculations at zero load as well
(Figs.\ \ref{fig:tf0} and \ref{fig:r0}). These results indicate that
for many $N$, $\Nd$, and $\Ld$ combinations $\tf$ is below the $0.01$~s
limit imposed by the experiments, even at
$\kb=20$~s$^{-1}$$\mu$M$^{-1}$. Also, the values of $r$ for
$\Ld=3.5$~nm are orders of magnitude larger than $1$, so with neck
linker docking backward steps are highly improbable even without any
rectification, in accordance with the experiments. Nevertheless, as $r$
can never be smaller than $1$ at zero load, the same rectification
mechanism that helps at the stall force is also able to make the
completion of backward steps rather unlikely in the unloaded situation.

\section{Conclusions}

As the results indicate kinesin under no load could march along MT in
the forward direction even without neck linker docking. Under the stall
force, however, neck linker docking seems to be crucial to forward
stepping, as for $\Nd=0$ (corresponding to no docking) the
characteristic forward binding time $\tf$ is way over the desired value
of $0.3$~s. If the binding rate constant $\kb$ near the MT is not
expected to be larger than its bulk value of $20$~s$^{-1}$$\mu$M$^{-1}$
by more than a factor of $3$, then we get very strict constraints for
the neck linker: it should contain at least $N=6$ Kuhn segments, and
dock all along the head with a horizontal projection of $\Ld=3.5$~nm.
Moreover, since the value of $r$ quickly decreases as $N$ is increased,
the best strategy for kinesin to walk forward upto a loading force of
$-7$~pN is to have a neck linker of about $N=6$ Kuhn segments
(corresponding to $13$-$14$ amino acids), with $\Nd=4$ or $5$ segments
capable of docking along the head. These values are very close to those
known from the structure of kinesin, suggesting that kinesin is a very
finely tuned motor protein whose parameters are set to live up to the
highest load possible. This coincidence not only validates our
calculations, but also justifies the relevance of neck linker docking
in the motility of kinesin.

\section{Appendix}

Here we derive the formula for $\rho_{N}^{0}(\vR)$, the three
dimensional probability density of the end-to-end distance of a FJC
with $N$ segments at zero loading force. Since $N$ can be quite small we
are not allowed to use the long FJC approximation of a polymer chain.

Let $\rho(\vR)$ be any three dimensional probability density with
spherical symmetry so that $\vR=0$ is the origin. Then the relationship
between $\rho(\vR)$ and its projection $\tilde\rho(x)$ to the $x$ axis
is:
\begin{equation}
\rho(\vR) =
 -\frac{1}{2 \pi |\vR|} \frac{\dd\tilde\rho(x)}{\dd x}\Bigg|_{x=|\vR|}
\label{eq:1d3d}
\end{equation}

Since the end of a rigid rod can only move on the surface of a sphere
the three dimensional density for an $N=1$ FJC is a Dirac delta
function on the surface of a sphere with radius $\LK$:
$\rho_{1}^{0}(\vR) = \delta(\LK-|\vR|)/(4\LK^2\pi)$. By substituting
$\rho_{1}^{0}(\vR)$ into Eq.\ (\ref{eq:1d3d}) and integrating it we get
for $\tilde\rho_{1}^{0}(x)$ a constant function with a value of
$1/(2\LK)$ and a support between $-\LK$ and $\LK$. After
autoconvoluting this function $N$ times to get $\tilde\rho_{N}^{0}(x)$,
and then converting it back to $\rho_{N}^{0}(\vR)$ with the help of
Eq.\ (\ref{eq:1d3d}), we arrive at the desired formula.

\section{Acknowledgments}

This work was supported by
the Hungarian Science Foundation (K60665)
and the Human Frontier Science Program (RGY62/2006).

\end{document}